\documentclass[12pt]{iopart}
\usepackage{graphicx}
\usepackage{color}
\usepackage{cite}
\usepackage{iopams}  
\begin{document}

\title{Effects of particle stiffness on the extensional rheology of model rod-like nanoparticle suspensions}

\author{Christian Lang,\textit{$^{a}$}$^{\dagger}$ Jan Hendricks,\textit{$^{b}$} Zhenkun Zhang,\textit{$^{c}$} Naveen K. Reddy,\textit{$^{d}$} Jonathan P. Rothstein,\textit{$^{e}$} M. Paul Lettinga,\textit{$^{a}$}$^{\dagger}$ Jan Vermant\textit{$^{f}$} and Christian Clasen\textit{$^{b}$}}

\address{\textit{$^{a}$~Institute of Complex Systems-3, Forschungszentrum J\"ulich, 52428 J\"ulich, Germany; Tel: 0049 2461 61 2149}}
\ead{c.lang@fz-juelich.de}
\address{\textit{$^{b}$~Department of Chemical Engineering, KU Leuven, B-3001 Leuven, Belgium}}
\address{\textit{$^{c}$~Key Laboratory of Functional Polymer Materials, Ministry of Education and Institute of Polymer Chemistry, Nankai University, Tianjin, 300071, P. R. China}}
\address{\textit{$^{d}$~Faculty of Industrial Engineering, Hasselt University, Martelarenlaan 42, 3500 Hasselt, Belgium and IMO, IMOMEC, Hasselt University, Wetenschapspark 1, 3590 Diepenbeek, Belgium}}
\address{\textit{$^{e}$~Mechanical and Industrial Engineering, University of Massachusetts, Amherst, MA 01003, USA}}
\address{\textit{$^{f}$~Department of Materials, ETH Z\"urich, 8093 Z\"urich, Switzerland }}
\vspace{10pt}

\begin{abstract}
The linear and nonlinear rheological behavior of two rod-like particle suspensions as a function of concentration is studied using small amplitude oscillatory shear, steady shear and capillary breakup extensional rheometry. The rod-like suspensions are composed of fd virus and its mutant fdY21M, which are perfectly monodisperse, with a length on the order of 900~nm. The particles are semiflexible yet differ in their persistence length. The effect of stiffness on the rheological behavior in both, shear and extensional flow, is investigated experimentally. The linear viscoelastic shear data is compared in detail with theoretical predictions for worm-like chains. The extensional properties are compared to Batchelor's theory, generalized for the shear thinning nature of the suspensions. Theoretical predictions agree well with the measured complex moduli at low concentrations as well as the nonlinear shear and elongational viscosities at high flow rates. The results in this work provide guidelines for enhancing the elongational viscosity based on  purely frictional effects in absence of strong normal forces which are characteristic for high molecular weight polymers.
\end{abstract}

\footnotetext{Also at: Laboratory for Soft Matter and Biophysics, KU Leuven, 3000, Leuven, Belgium}

%
\vspace{2pc}
\maketitle
%
%

\section{Introduction}
Non-spherical particles are more commonly found in nature than their spherical counterparts, and the chemical composition ranges from inorganic nanoparticles \cite{Kingery1976, Hradil2003, Bailey2007} to biological filaments \cite{Stanley1941, Elson1988, Cabeen2005}. There are many advantages for using non-spherical and in particular rod-like nanoparticles in industrial applications. First, stable rod-like particles are well suited as rheology modifiers, enhancing the shear thinning \cite{Solomon2010} and increasing the extensional viscosity \cite{Mewis1974} without imparting significant normal stress differences to the solution, as is the case for polymers, and for usage in spinning or spraying applications. Second, for attractive systems the percolation treshold is lowered \cite{Schilling2007, Zhang2009, Huang2009, Reddy2012}, which is exploited for example in sag control agents. Non spherical suspensions also have a fascinating phase behavior \cite{Grelet2003, Dogic2006}. Furthermore, rod-like nanoparticles are commonly used to enhance mechanical, electrical and optical properties in suspensions \cite{Kyrylyuk2008, Schultz2009} or composites \cite{Srivastava2003, Hobbie2009}, as well as in coatings \cite{Kheirandish2008} and structured consumer products \cite{Maynard2006}. Similarly, their mechanical properties play an important role in biological systems, for example as a constituent of the cytoskeleton in cells \cite{Elson1988, Lin2007}.\par
The rheology of rod-like nanoparticle suspensions was reviewed by Solomon \cite{Solomon2010} who identified four physiochemical properties that control their behavior: interactions between particles, aspect ratio, persistence length, and number density. Most systematic studies have been carried out for rod-like nanoparticle suspensions with relatively low aspect ratios $L/d<20$ \cite{Zrinyi1993, Solomon1998, Mohraz2006}. Additionally, many large aspect ratio rod-like particle suspensions, e.g. carbon nanotubes, glass fibers or actin, are typically polydisperse. The most extensively studied rod-like particles are carbon nanotubes and actin filaments, of which the material functions in shear and oscillatory flow have been determined. Intrinsic viscosities from shear rheology at low P\'eclet numbers are successfully used in determining the length of carbon nanotubes using the Kirkwood-Batchelor theory \cite{Davis2004, Nicholas2007, Marceau2009, Kirkwood1956, Batchelor1970}, which is in agreement with atomic force microscopy (AFM) results. The viscoelastic behavior of carbon nanotubes has been shown to arise predominantly from aggregate formation, unlike actin, where the viscoelastic behavior is additionally effected by stretching and bending \cite{MacKintosh1995, Hough2004, Gardel2004}. For extensional rheology of rigid rod-like particles, the theories of Batchelor \cite{Batchelor1971}, and Shaqfeh \cite{Shaqfeh1990} have shown to be in quantitative agreement with experimental results for dilute fiber suspensions \cite{Mewis1974, Pittmann1990, Ma2008}. However, these high aspect ratio rod-like particles have small persistence lengths, are polydisperse, and lack longterm colloidal stability \cite{Joly2002, Huang2006}. In the present work, we pursue an experimental study of both the shear and extensional  flow properties on a model system of monodisperse virus nanoparticles, where the stiffness can be modified by genetic mutation. In addition, due to surface charges, these colloids have kinetic stability.\par
\begin{figure}
  \centerline{\includegraphics[scale=0.7]{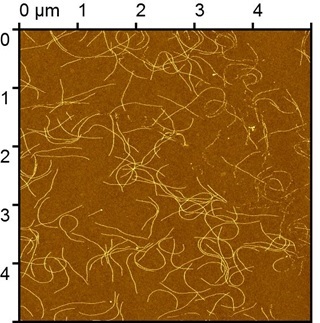}}
  \caption{Atomic force microscopy picture of fd virus dispersed on a Poly(allylamine)/Silicon substrate.}
\label{fig1}
\end{figure}

In the present study, we use the bacteriophages fd (wild type) and its mutant fdY21M \cite{Sambrook2001, Barry2009}. These virus particles are monodisperse and are in the colloidal domain ($<1~\mu$m), see Fig.~\ref{fig1}. For the wild type fd virus, the dimensions are $L=880$~nm and $d=6.6$~nm with a persistence length  $L_{p}=2800\pm700$~nm \cite{Dogic2006}, while the stiffer mutant virus fdY21M is characterized by a slightly larger $L=910$~nm, $d=6.6$~nm and a 3.5 times larger persistence length $L_{p}=9100\pm1600$~nm \cite{Barry2009}. The viruses are inherently negatively charged at neutral pH \cite{Grelet2003} and therefore stable in buffer solutions. The surface of both viruses consists of proteins, helically ordered along the core of circular DNA \cite{ Zhang2010}. For the mutant virus fdY21M, the 21$^{st}$ amino group of the capsid protein (g8p) is changed from tyrosine (Y) to methionine (M), causing a drastic change in stiffness because the larger methionine introduces a tighter packing of the virus coating. Since both bacteriophages are grown from a single colony of molecular clones of E.-coli bacteria, they are monodisperse by nature.\par
For suspensions of virus particles, only limited rheological data is available. Steady shear rheology was reported by Lang et al. \cite{Lang2016}, while other studies focus on the scaling of the material functions with particle concentration \cite{Graf1993, Schmidt2000}, salt concentration \cite{Gomez2012}, and the detection of the suspension to gel transition \cite{Zhang2009, Huang2009}. Close to the I-N transition, these systems were also interesting model systems to investigate shear banding phenomena \cite{Kang2006, Kang2008, Dhont2010}. It has been observed in literature that due to the principle differences in the deformation profiles of shear and extensional flow, the evolution of rheological properties of microstructured fluids are drastically different \cite{Smith1998, Smith1999}, with a strongly enhanced elongational viscosity\cite{Prudhomme1994, Cathey1988}. The Trouton ratio, $T_{r}=\eta_{e}/\eta$, defined as the ratio between elongational viscosity, $\eta_{e}$, and shear viscosity, $\eta$, at the same flow strength is a good way to characterize this. For a Newtonian fluid $T_{r}=3$, while this should not hold for suspensions of rod-like particles.
According to our knowledge, no experiments have been conducted to measure the extensional properties of suspensions made from model high aspect ratio rod-like particles, in particular evaluating the effect of persistence length. Therefore, the goal of this paper is to identify the differences between shear and uniaxial extensional flow behavior, illustrating also the influence of particle stiffness.\par 
The paper is organized as follows. First, the relevant theories are briefly introduced followed by materials and methods. Then, we compare the data from linear and nonlinear rheological measurements in shear as well as elongational flow to the theoretical predictions.

\section{Theory}
Theoretical studies relating the structural properties of rod-like particles in a Newtonian medium to the their rheological behavior began with the work of Jeffery \cite{Jeffery1922}, who described the motion of ellipsoidal particles in a simple shear flow in the dilute concentration regime $\varphi<\varphi^\ast$, where $\varphi$ is the volume fraction and $\varphi^\ast\sim1/dL^2$ is the overlap volume fraction, with $L$ and $d$ the length and thickness of the particles. To extend the analysis into the semi-dilute concentration regime, $\varphi^\ast<\varphi<\varphi_{IN}$, where $\varphi_{IN}$ is the isotropic nematic spinodal point, information on the effect of rotational diffusion due to Brownian motion becomes necessary. The rotational diffusion for very slender rod-like particles was first derived by Broersma \cite{Broersma1960}:
\begin{equation}
\label{diffCoeff}
D_{r}=\frac{3k_BT\ln{L/d}}{\pi\eta_{s}L^3}~~,
\end{equation}
where $k_BT$ is the unit energy and $\eta_{s}$ is the solvent viscosity. The relative strength of the applied force field to Brownian motion is then given by the rotational P\'eclet number
\begin{equation}
\label{Peclet}
Pe=\frac{\dot{\gamma}}{D_{r}}~~,
\end{equation}
where $\dot\gamma$ is the shear rate.\par
\begin{figure}
  \centerline{\includegraphics[scale=0.7]{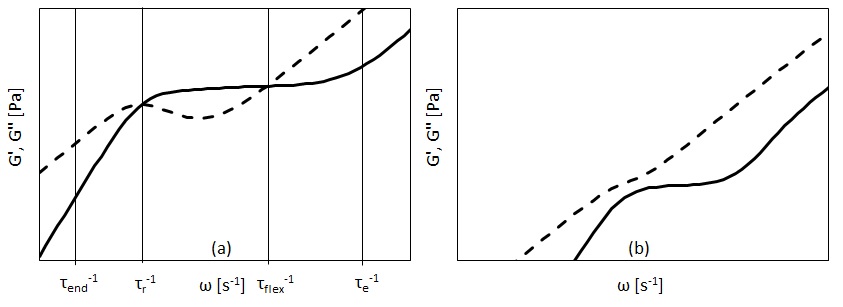}}
  \caption{Storage and loss modulus vs. angular frequency from Morse theory for $L_{p}\gg L$ in the semi-dilute (a) and dilute case (b). Vertical lines in (a) indicate the location of the most important relaxation times.}
\label{fig2}
\end{figure}

In the small P\'eclet number regime, the prediction of rheological properties for semidilute rod-like suspensions is given by the theories of Kirkwood, Batchelor, Doi \cite{Kirkwood1956, Batchelor1970, Doi1977b}, and later Morse \cite{Morse1998a, Morse1998b}.In addition to the rotational relaxation time of rods, $\tau_{r}=1/D_{r}$, for semiflexible rods, a wormlike chain model can be used to capture the additional relaxation mechanisms, using the steric geometry and persistence length of the rods as well as the the solvent viscosity and concentration of the suspension as key parameters. In particular, due to entanglement effects between particles in the semidilute regime, a long time rotational diffusion, $\tau_{end}=1/\langle D_{r}\rangle$, is introduced that relates to the dilute $D_{r}$ via: 
\begin{equation}
\label{meanRotDiff}
\langle D_{r}\rangle\sim D_{r}(\nu L^3)^{-2}~~,
\end{equation}
with particle number density $\nu$, and which is much longer than $\tau_{r}$ \cite{Doi1977}. At shorter times, inter-particle stresses can relax via particle undulations with a relaxation time, $\tau_{flex}$, and the shortest relaxation time, $\tau_{e}$, is that of a single undulation mode \cite{Shankar2002}. Using small amplitude oscillatory shear flow, one can measure $\tau_{r}$ and $\tau_{flex}$, and for concentrations close to $\varphi_{IN}$ also $\tau_{e}$ can be estimated. The relaxation time spectrum of rod-like colloidal suspensions, for which $ L_{p}\gg L$, in the semidilute concentration regime is sketched in Fig.~\ref{fig2}(a) and the relaxation times are indicated. For dilute rod-like suspensions, the theory predicts essentially the same curve form for the storage modulus, as shown in Fig.~\ref{fig2}(b), but due to a monotonic increase of the loss modulus, the moduli curves do not cross each other.\par
The zero shear viscosity $\eta_0$ observed under steady state shear flow for $Pe\ll1$ was theoretically predicted by Doi \cite{Doi1977b}, and  \cite{Shaqfeh1990}, and follow-up theories including hydrodynamics, e.g. by Djalili et al. \cite{Djalili2005}. All these theories include at least one fitting parameter and generally underpredict $\eta_0$. Since the zero shear viscosity of rod-like suspensions is hard to measure, as has been discussed elsewhere \cite{Lang2016}, we cannot compare the theories with measurements.

For $Pe\gg1$, the flow is dominated by hydrodynamic forces. In this regime, the effect of both, aspect ratio and P\'eclet number, on the orientational ordering of rod-like particles in flow was given by Hincha and Leal \cite{Hinch1972} (HL). In the regime $1\ll Pe\ll(L/d+d/L)^3$, they predict an intrinsic viscosity of:
\begin{equation}
\label{HL}
[\eta]=\frac{\eta-\eta_{s}}{\eta_{s}\varphi}=Pe^{-1/3}\frac{0.5(L/d)^2}{\ln{L/d}}~~.
\end{equation}
We test this assertion using steady shear rheology. At high P\'eclet numbers, the shear viscosity should approach $\eta_{s}$ independent of the P\'eclet number.\par
Theoretical predictions of the extensional viscosity $\eta_{e}$ of rods in the high P\'eclet number regime were first derived by Batchelor \cite{Batchelor1971} for dilute suspensions and later by Shaqfeh \cite{Shaqfeh1990}. Batchelor's prediction for the intrinsic elongational viscosity reads:
\begin{equation}
\label{Batchelor}
[\eta_{e}]=\frac{\eta_{e}-3\eta_{s}}{3\eta_{s}\varphi}=\frac{4(L/d)^2}{9\ln{\pi/\varphi}}~~,
\end{equation}
while the Shaqfeh-Fredrickson (SF) analysis adds additional terms due to finite size effects and intermediate range hydrodynamic interactions:
\begin{equation}
\label{SF}
[\eta_{e}]=\frac{4(L/d)^2}{9(\ln{1/\varphi}+\ln\ln{1/\varphi})+0.1518}~~.
\end{equation}
We test both theories using data from capillary breakup extensional rheometry (CaBER). One of the important parameters that is useful in processes dominated by extensional flow is the Trouton ratio. In order to calculate appropriate Trouton ratios, we have to compare $\eta$ and $\eta_{e}$ at the same flow strength, given by the square root of the second invariant of the rate of deformation tensor $\sqrt{|II_{2D}|}=\dot\gamma=\sqrt{3}\dot\epsilon$, where $\dot\epsilon$ is the extension rate. 
 
\section{Experimental}
\subsection{Materials}

In order to make fd virus suspensions more suitable as a model system for rheological investigation, the virus is suspended in glycerol-water mixtures in order to slow down the particle dynamics and possible sedimentation effects, and to obtain an optical matching of the refractive indices that lowers the Van der Waals attractions. In this manner, both macroscopic and microscopic long-term colloidal stability is ensured, thus allowing for long time measurements as well as low frequency and low rate rheological experiments.\par
Anhydrous glycerol ($\geq$~99.5~\%) and Trizma base were obtained from Sigma Aldrich and used as received without further purification. Mixing 20~mM water-based Tris buffer with 90~mM NaCl resulted in an ionic strength of 100~mM. This buffer solution was mixed with glycerol to obtain two buffered water/glycerol mixtures of 86.01 and 86.03~wt\% glycerol with an ionic strength of 16~mM.\par 
Fd virus and fdY21M were grown and purified following standard biological protocols \cite{Sambrook2001}, using the XL-1 blue strain of E.-coli as a host bacterium. The yield from 6~L of growth medium is about 90~mg of virus. The virus particles were cleaned by multiple centrifugation steps at 3.6x$10^4$~g for 2~h  each and ultra-centrifugation at 1.1x$10^5$~g for 12~h. In the final step of purification, virus particles were redispersed in a small volume of 100~mM Tris buffer to produce a highly viscous slurry. Under continuous stirring, the slurry was mixed with anhydrous glycerol until a weight fraction of 86.03~m\% glycerol for the fd medium and 86.01~m\% for the fdY21M medium was reached. Further dilution was accomplished by mixing the mother solutions with glycerol-Tris buffer mixtures of the same compositions. The glycerol/water ratio was chosen to minimize water uptake from the surrounding air.

\subsection{Methods}
\begin{figure}
  \centerline{\includegraphics[scale=0.6]{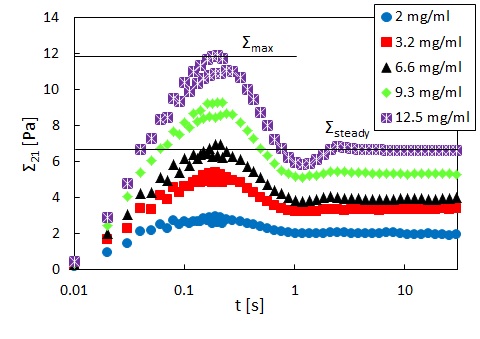}}
  \caption{Shear stress as a function of time for different concentrations of fd virus at a shear rate of 10~s$^{-1}$. The horizontal lines mark the stress level at the overshoot $\Sigma_{max}$ and in the steady state $\Sigma_{steady}$ for a concentration of 12.5~mg/ml.}
\label{fig3_1}
\end{figure}

The rheological properties of the virus suspensions in glycerol-Tris buffer were investigated using a strain controlled ARES LS rheometer (TA Instruments, New Castle, England). All rheometry experiments were carried out at 22$^o$C with a cone-plate geometry of 25~mm diameter and a cone angle of 2.5$^o$, using a solvent trap. The samples were pre-sheared at 100~s$^{-1}$ for 5~min with a subsequent resting period of 100~min to avoid artifacts from sample loading and for equilibration of structure and temperature. In order to assess the lower time limit for steady shear experiments, the transient viscosity was recorded over time and the steady state viscosity was obtained from the final plateau value, see Fig.~\ref{fig3_1}. Using this protocol, steady state viscosity data was obtained for shear rates between 0.01~s$^{-1}$ and 1000~s$^{-1}$. All flow curves as well as oscillatory experiments were determined using samples from one batch and repeated at least three times to ensure reproducibility. For dynamic measurements, strain amplitude sweeps were carried out prior to frequency sweeps to access the linear viscoelastic regime. Frequency sweeps were carried out within a frequency range for which restructuring time scales, $\tau_{r}$, were longer than the measurement times. The restructuring half time of 108.9~s was obtained from stress relaxation measurements comparable to the relaxation of the Kerr effect \cite{Doi1977}, which sets the lower limit for the frequency to 0.0092~$s^{-1}$.\par

\begin{figure}
  \centerline{\includegraphics[scale=0.6]{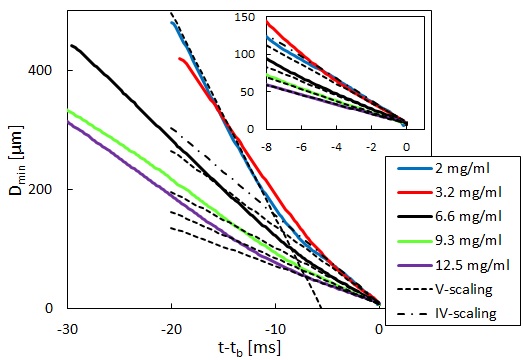}}
  \caption{Filament diameter versus time for different concentrations of fd virus. Dashed lines indicate the visco-capillary (V)  scaling and dash-dotted lines indicate the inertia-viscous (IV) scaling. Inset: Filament diameter versus time in the terminal region.}
\label{fig3}
\end{figure}

Uniaxial extensional rheometry was performed in a Haake CaBER-1 extensional rheometer (Thermo Haake GmbH, Karlsruhe, Germany), where the drive unit was used to control position and separation velocities of two circular parallel plates with a diameter $D_p=4$~mm. The evolution of the thinning fluid filament between the two plates was recorded by a high speed camera \cite{Matheus2015} (Photron Fastcam SA-2, Photron, San Diego CA, USA). The video images were subsequently analyzed by digital image processing implemented in Matlab in order to determine the full filament profile along with the position and dimension of the minimum filament radius $R_{min}$ \cite{Matheus2015}. To control the liquid filament thinning, the fluid was extended by rapidly stretching the fluid from an initial plate separation of 2~mm to a plate separation of 6~mm. Extensional viscosities were calculated from fitting the minimum diameter evolution close to breakup using a similarity solution by Papageorgiou \cite{Papageorgiou1995}:
\begin{equation}
\label{rmin}
R_{min}(t)=R_1-\frac{(2X-1)\sigma}{6\eta_{e}}t~~,
\end{equation}
where $X=0.7127$ is a correction factor accounting for deviation from a perfect cylindrical filament shape \cite{Papageorgiou1995, McKinley2000}, $\sigma$ is the surface tension of the suspension, and $R_1$ is the filament radius at the beginning of the terminal time regime. In the final stage of the breakup process, for low viscous fluids a universal similarity solution with X=0.5912 is eventually reached, which was determined by Eggers \cite{Eggers1993, Eggers1997}, and the thinning dynamics are dominated by visco-inertial forces. This visco-inertial regime is only reached for the lowest concentrations in our case. The extension rate was calculated as $\dot{\epsilon}(t)=(-2/D_{min})(dD_{min}(t)/dt)$, where $D_{min}$ is the minimum filament diameter before imposing the stretch. Each experiment was repeated at least five times to ensure reproducibility. In Fig.~\ref{fig3}, $D_{min}$ is shown as a function of time, with $t_{b}$ being the time of filament breakup. Each line represents one typical experiment. The dashed and dash-dotted lines are fits of the similarity solution used for the viscosity calculation in the viscous case (V-scaling) as well as the inertial-viscous case (IV-scaling). At early times and large minimum diameter, the filament does not fulfill the slenderness assumption required for the application of eq.~\ref{rmin} and gravitational forces cannot be neglected\cite{Campo2010}. With decreasing diameter, a visco-capillary balance (V-scaling) is approached which is eventually followed by an inertia-viscous balance (IV-scaling). With increasing viscosity, the transition between the V-scaling and the IV-scaling moves to lower filament diameters and appears at diameters below the optical resolution limit for concentrations above 2~mg/ml. \par 
The surface tension of each system was determined using the pendant drop method. A value of $\sigma=63\pm1$~mN/m was obtained for all suspensions under investigation, independent of the particle concentration.

\section{Results}

\subsection{Small amplitude oscillatory shear}
\begin{figure}
  \centerline{\includegraphics[scale=0.6]{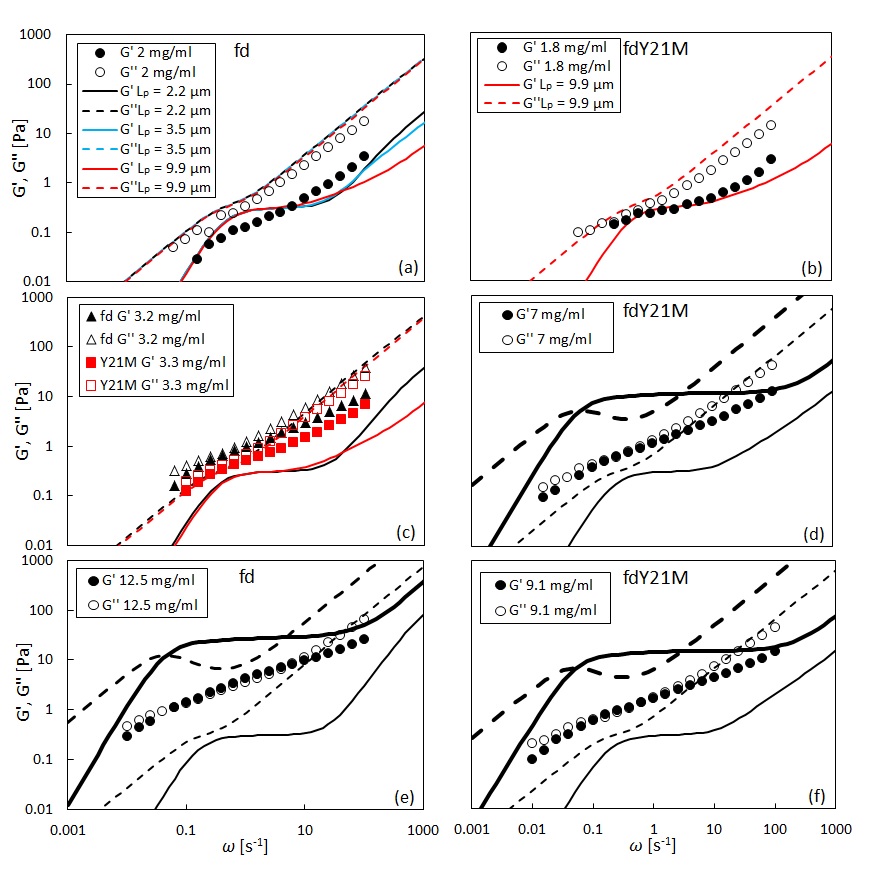}}
  \caption{Complex moduli $G'$ (full symbols, full lines) and $G''$ (open symbols, dashed lines) as a function of frequency for different concentrations of (a), (e): fd; (b), (d), (f): fdY21M and (c) both viruses. The lines are predictions from Morse theory in the dilute (thin lines) and concentrated (thick lines) regime. Numbers in the caption of (a) and (b) indicate different theoretical persistence lengths. For the rest of the figure persistence lengths of 2.2~$\mu$m for fd and 9.9~$\mu$m for fdY21M are used.}
\label{fig4}
\end{figure}

\begin{figure}
  \centerline{\includegraphics[scale=0.7]{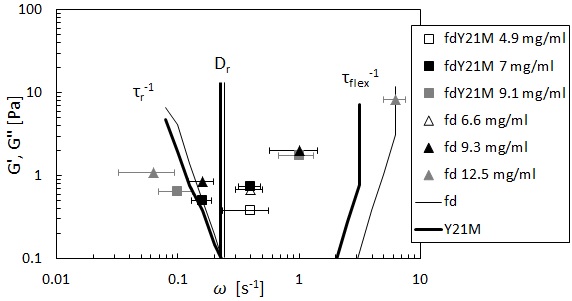}}
  \caption{Touching (open symbols) and crossover points (closed symbols) of moduli compared to Morse theory for fd (thin line) and fdY21M (thick line). The straight lines show $D_{r}$ computed from equation~\ref{diffCoeff}. The lower frequency crossover points mark the inverse rotational relaxation time $\tau_{r}^{-1}$ and the higher frequency cross over points are the inverse undulation time $\tau_{flex}^{-1}$.}
\label{fig5}
\end{figure}

In the linear viscoelastic regime, the relaxation times of fd virus suspensions were probed using small amplitude oscillatory shear and evaluate the dynamic moduli, $G'(\omega)$ and $G''(\omega)$, as a function of the angular frequency, $\omega$. Fig.~\ref{fig4} shows the linear shear rheology of the flexible fd and stiff fdY21M. For the low nanoparticle concentrations (1~mg/ml<c<3~mg/ml) in Figs.~\ref{fig4}(a) and (b), both virus suspensions show liquid like behavior with $G''>G'$ at all experimentally accessible frequencies. Despite the obvious differences between the moduli of the two species at low concentrations, the dynamic moduli curves are of similar form at intermediate concentrations. Here, both of the moduli curves touch each other, having equal values at a certain frequency, as indicated in Fig.~\ref{fig4}(c) as well as by the open symbols in Fig.~\ref{fig5}. At even higher concentrations (c>4~mg/ml), an elastic plateau region emerges, in which $G'>G''$, producing two crossover points, indicated in Figs.~\ref{fig4}(e) and (f) as well as by the closed symbols in Fig.~\ref{fig5}. The two crossover points mark the relaxation times of rotational self diffusion and chain undulation, $\tau_{r}>\tau_{flex}$. The inverse relaxation times, $1/\tau_{r}$, represent a measure of rotational self diffusion and are relatively low due to the high viscosity suspending medium. Equation~\ref{diffCoeff} gives $D_{r}=1/\tau_{r}=0.22~s^{-1}$ for fdY21M and $D_{r}=0.23~s^{-1}$ for fd in the dilute limit. While we find lower frequency crossover points similar to those predicted by equation~\ref{meanRotDiff}, located between $0.06~s^{-1}<1/\tau_{r}<0.2~s^{-1}$ for both rods, the reciprocal relaxation time of chain undulation lies between $0.4~s^{-1}<1/\tau_{flex}<1~s^{-1}$ for the stiff rod fdY21M and $0.4~s^{-1}<1/\tau_{flex}<6.3~s^{-1}$ for the flexible fd. A broadening of the elastic region with increasing concentration can be clearly seen in Fig.~\ref{fig5}, where the closed symbols represent the low and high frequency boundaries of this region for different concentrations compared to the predictions by Morse theory. This broadening between crossover points in the direction of higher frequencies for fd virus compared to the fdY21M virus indicates the importance of stress relaxation due to chain undulation for more flexible rod-like particles. From the values for $\tau_{e}\sim L_{e}^4/D_{r}L_{p}$, located at the onset of equal slope of both moduli, it is possible to estimate the average tube length $L_{e}\sim L_{p}(\rho L^2)^{-2/5}$, where $\rho$ is the contour length per unit volume, of both systems \cite{Morse1998b}. The identifiable values of $1/\tau_{e}$ seem to be concentration independent. We find $L_{e}\approx0.3~\mu$m for fd virus and $L_{e}\approx0.9~\mu$m for fdY21M, thereby, roughly confirming the difference in persistence length between the two systems.

The theoretical prediction by Morse \cite{Morse1998b} agrees qualitatively with the experimental moduli for flexible fd whereas for fdY21M it agrees both qualitatively and quantitaively at low concentrations, as shown in Figs.~\ref{fig4}(a) and \ref{fig4}(b). Interestingly, the predictions for the wild type fd are not in agreement with the experiments. Since the agreement with the storage modulus of fdY21M is better, we investigated whether a change in persistence length for the Morse prediction of fd virus leads to a more quantitative agreement. Fig.~\ref{fig4}(a) shows a comparison of the experiment with three different theoretical curves obtained for three persistence lengths up to and including the persistence length of fdY21M. The theoretical predictions for different stiffnesses are indistinguishable for lower frequencies and differ only in the high frequency regime. Therefore, we conclude that Morse theory gives a better estimate of the complex moduli of stiffer rods as compared to more flexible ones. Another possibility for the discrepancies between theory and experiment could be the use of a wrong effective thickness in the definition of the number density used in Morse theory, resulting from the unknown zeta potential of our solvent. We applied a large range of different particle thicknesses, from 6 to 25~nm, but the sensitivity of the predictions to the thickness is limited.\par
Morse theory seems to overestimate the magnitude of the intrinsic moduli at higher concentrations and predicts a more complex curve shape than observed in our experiments, see Figs.~\ref{fig4}(e), (d), and (f), whereas experiments on significantly longer rods show indeed the predicted curve shape \cite{Schmidt2000a}.  From Figs.~\ref{fig4}(e) and (f) as well as from Fig.~\ref{fig5}, it can be observed that the theory predicts a decrease of $1/\tau_{flex}$ with increasing stiffness. As indicated by the theoretical lines in Fig.~\ref{fig5}, however, the broadening of the elastic region observed in our experiments seems to be more drastic than the theoretical prediction.

At low frequencies, both systems should show a strongly reduced rotational diffusion $\langle D_{r}\rangle\approx0.012~s^{-1}$ due to entanglement effects. Despite the highly viscous buffer solution, we are not able to clearly identify the slope of the moduli curves in the low frequency regime or the diffusivity in the entanglement affected long time regime. Therefore, zero shear viscosities of the suspensions, $\eta_0=\lim_{\omega\to0}G''(\omega)/\omega$, are not measurable from the oscillatory data. 

\begin{figure}
	\centerline{\includegraphics[scale=0.7]{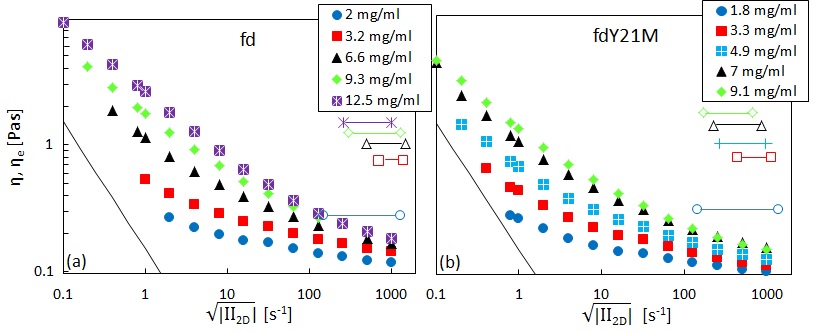}}
	\caption{Shear (full symbols) and elongational viscosity (open symbols), as a function of the square root of the second invariant of the rate of deformation tensor for (a): fd and (b): fd Y21M at different concentrations. The thin line marks the lower torque limit of the rheometer.}
	\label{fig7}
\end{figure}

\subsection{Steady shear viscosity}
\begin{figure}
	\centerline{\includegraphics[scale=0.7]{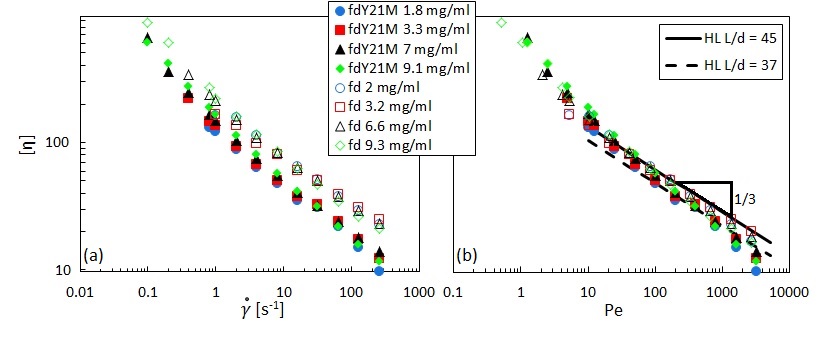}}
	\caption{(a): Intrinsic viscosity as a function of shear rate for different concentrations. (b): Intrinsic viscosity as a function of P\'eclet number as defined by equation~\ref{Peclet}. The line shows a prediction by HL theory.}
	\label{fig6}
\end{figure}

Fig.~\ref{fig7} displays the shear viscosity of fd (full symbols) and fdY21M (open symbols) as a function of shear rate $\sqrt{|II_{2D}|}=\dot{\gamma}$ for different concentrations. The shear viscosity of the flexible virus, fd, given in Fig.~\ref{fig7}(a), is higher than the viscosity of the stiff virus, fdY21M, shown in Fig.~\ref{fig7}(b), for all shear rates. The highly concentrated samples display the strongest shear thinning behavior in the low shear rate region compared to other concentrations. 
Fig.~\ref{fig6}(a) shows the shear rate dependent intrinsic viscosity of fd and fdY21M following equation~\ref{HL}, calculated from the data in Fig.~\ref{fig7}. Using the intrinsic viscosity instead of the shear viscosity itself brings all flow curves of one type of rod at different concentrations to a single master curve. The intrinsic viscosity of the more flexible fd virus is still higher than the intrinsic viscosity of the stiffer rod fdY21M for the complete measurable shear rate range. In Fig.~\ref{fig6}(b), the shear rate is scaled by the rotational diffusion given in equation~\ref{diffCoeff}, bringing the two measured curves closer together. In the high P\'eclet number regime, the data is in good agreement with the theoretical prediction by Hinch and Leal \cite{Hinch1972}, combining equation~\ref{HL} with a fitted aspect ratio of 45 for fd and 37 for fdY21M. This points in the direction of hydrodynamic interactions dominating the high shear rate regime. Using these effective aspect ratios, the estimated hydrodynamic radii of the two rods, given their well-known length, are 19.6~nm for fd and 24.6~nm for fdY21M, which is significantly larger than the bare diameter of 6.6~nm. This can be attributed to hydrodynamic interactions that decay over a distance of $\sim 2d$ for aligned rods\cite{Lettinga2010}, whilst the glycerol also might form a solvation layer. 

\subsection{Uniaxial extension}
The apparent extensional viscosity of rod-like particles was measured using capillary breakup extensional rheometry. In Fig.~\ref{fig7}, we show a comparison between the shear viscosity and the extensional viscosity, plotted as a function of the square root of the second invariant of the rate of deformation tensor $\sqrt{|II_{2D}|}=\dot{\gamma}=\sqrt{3}\dot{\epsilon}$. The range of extension rates over which constant extensional viscosities are reported is a consequence of the type of experiment. The decreasing filament diameter in Fig.~\ref{fig3} causes a continuously increasing capillary pressure and thus for a Newtonian liquid an increasing extension rate. It is seen that the elongational viscosity is generally higher than the shear viscosity.   

\begin{figure}
	\centerline{\includegraphics[scale=0.6]{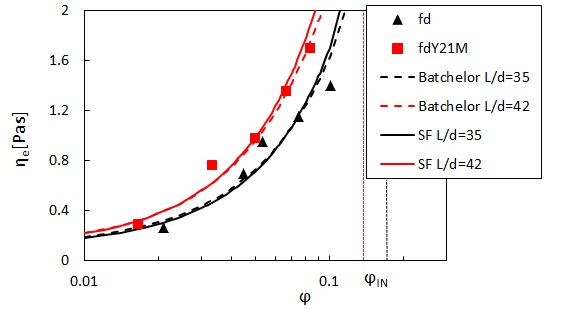}}
	\caption{Elongational viscosity of fd (black triangles) and fdY21M (red squares) as a function of concentration. The curves are predictions from Shaqfeh and Fredrickson (SF) and Batchelor theory, using the indicated aspect ratios. The vertical lines mark the position of $\varphi_{IN}$ for both systems.}
	\label{fig8}
\end{figure}

For Newtonian liquids, the Trouton ratio is $\eta_{e}/\eta=3$, independent of extensional rate and concentration. In rod-like colloidal suspensions, $\eta_{e}$  is predicted to be independent of $Pe$ in the limit of high P\'eclet numbers, however, it depends on concentration, as given in equations~\ref{Batchelor} and \ref{SF}. In Fig.~\ref{fig8}, the two theoretical curves are fitted to the measurement of the elongational viscosity. The theoretical prediction agrees with the measured data, if hydrodynamic aspect ratios of 35 for fd, and 42 for fdY21M are used as fitting parameters, or effective hydrodynamic thicknesses of $d=25.1$~nm for fd, and 21.7~nm for fdY21M. 
Thus, the hydrodynamic radius of the stiff rods under extensional flow  is \textit{larger} than that of the flexible rods, opposite of what was obtained from fitting the steady shear flow curves at high shear rates.

\begin{figure}
	\centerline{\includegraphics[scale=0.7]{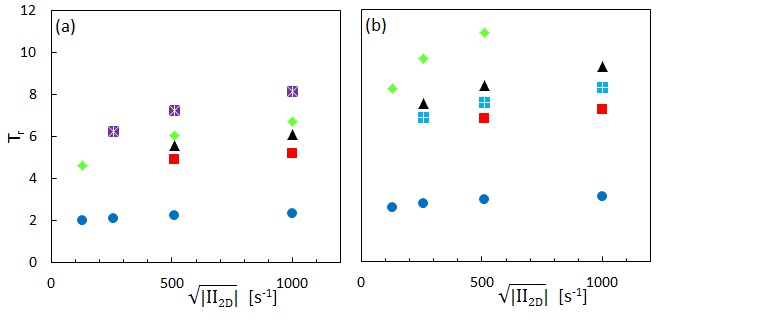}}
	\caption{Trouton ratio as a function of the square root of the second invariant of the rate of deformation tensor for different concentrations of (a) fd and (b) fdY21M. Legend, see figure~\ref{fig7}.}
	\label{fig9}
\end{figure}

Accordingly, we observe \textit{higher} Trouton ratios for the stiffer virus fdY21M compared to the more flexible virus fd, as can be seen in Fig.~\ref{fig9}, where the Trouton ratio of both systems is plotted as a function of $\sqrt{|II_{2D}|}$ for different concentrations. Again, this is markedly different from the behavior in steady shear flow where the viscosity of the stiff fdY21M is \textit{lower} than the values obtained for the more flexible virus fd.

\begin{figure}
	\centerline{\includegraphics[scale=0.6]{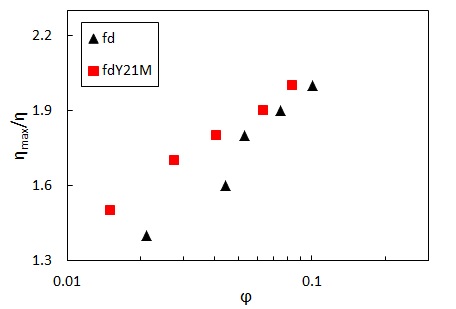}}
	\caption{Overshoot viscosity divided by steady state viscosity as a function of concentration at $\dot\gamma=10$~s$^{-1}$.}
	\label{fig10}
\end{figure}
\section{Discussion and Conclusions}

In this paper, we investigated the shear and extensional rheological properties of ideal stiff and more flexible rods, using rod-like viruses with similar aspect ratio ($L/d=120$) and different stiffness, respectively, $L=L_{p}/10$ and $L=L_{p}/3$. 
We find that at high rates the shear and extensional viscosity can be described by purely hydrodynamic theories for both systems: the shear viscosity displays the $1/3$ dependence on shear rate predicted by Eq. \ref{HL}, while the volume fraction dependence of the extensional viscosity is well described by Eq. \ref{SF}. This is in strong contrast to measurements on carbon nanotubes, a polydisperse system,  for which the concentration depence of the extensional viscosity is much stronger\cite{Darsono2012}.\par
For interpreting the results, we make use of the fact that the length of the systems is very well defined, so that the only fit parameter for both, extensional and shear viscosity, is the effective hydrodynamic radius. For the stiff rods, we find roughly the same value for the effective hydrodynamic radius from the two completely different types of experiments. Such as reported in the literature \cite{Lettinga2010}, the hydrodynamic radius exceeds the solvation layer of the particles.\par
We also find, however, that stiffness of the rods plays an important role. The effect of stiffness is of opposite character for both types of deformations.\par
The shear viscosity of the more flexible rods at high shear rates is \textit{higher} than for the stiff fdY21M. 
This could be due to the fact that flexible rods have the tendency to form complex structures such as hairpins\cite{Kirchenbuechler2014}, causing steric interactions, which can be neglected for stiff rods, as was done in both theories. This argument is not valid for extensional flow as in this case the more flexible virus should just stretch and behave very similar to the stiff virus. Still, the extensional viscosity is significantly \textit{smaller} for the more flexible virus. We do not have an explanation for this observation, but we do note that this is also the case for the viscosity at the point of stress overshoot to the steady state value, see Fig. \ref{fig10}, although in this case the difference decreases with increasing concentration.\par
So far, we discussed the effect of flexibility for very high rates, but the difference in flexibility between fd and fdY21M is also reflected in the linear dynamic data. Indeed, comparing $\tau_{flex}$ for both systems at high concentrations, as displayed in Fig. \ref{fig4}(d) and (e), we observe that this quantity is a factor 10 higher for the stiff system as compared to the flexible one. Moreover, $\tau_{r}$ of the flexible system compared to $\tau_{r}$ at the same volume fracton of the stiff rods is smaller, since the purely geometric hindrence of the flexible system is reduced, see Fig.~\ref{fig7}.\par
In conclusion, this work suggests that stiffness is a particularly important parameter in controlling the extensional viscosity of formulations containing rod-like particles. In order to support these suggestions, one would need to measure ordering at high shear rates, for example by performing Rheo-SANS measurements.

\section*{Conflicts of interest}
There are no conflicts to declare.

\ack
The authors would like to acknowledge David C. Morse for the fruitful discussions and O. Deschaume for performing AFM measurements. This study was funded by the European Union within the Horizon 2020 project under the DiStruc Marie Sk\l{}odowska Curie innovative training network; grant agreement no. 641839 as well as under the People Programme (Marie Curie Actions) of the European Union$'$s Seventh Framework Programme (FP7/2007-2013) under REA grant agreement no. 607937 (SUPOLEN), the Research Foundation Flanders (FWO, grant no. G077916N), and a senior fellowship (SF/14/018, KU Leuven Internal Funds).


\bibliography{bib1}
\bibliographystyle{plain}

\end{document}